\documentstyle[aps,prl,multicol,epsf]{revtex}

\newcommand{\bleq}{\ifpreprintsty
                   \else
                   \end{multicols}\vspace*{-3.5ex}{\tiny
                   \noindent\begin{tabular}[t]{c|}
                   \parbox{0.493\hsize}{~} \\ \hline \end{tabular}}
                   \fi}
\newcommand{\eleq}{\ifpreprintsty
                   \else
                   {\tiny\hspace*{\fill}\begin{tabular}[t]{|c}\hline
                    \parbox{0.49\hsize}{~} \\
                    \end{tabular}}\vspace*{-2.5ex}\begin{multicols}{2}
                    \fi}
\newcommand{\bcols}{\ifpreprintsty\else\begin{multicols}{2}\fi}
\newcommand{\ecols}{\ifpreprintsty\else\end{multicols}\fi}

\begin{document}

\title{Distribution of parametric conductance derivatives of a quantum dot}

\draft

\author{P. W. Brouwer,$^a$ S. A. van Langen,$^a$
  K. M. Frahm,$^a$\thanks{Present address: Laboratoire de Physique
  Quantique, Universit\'e Paul Sabatier, 31062 Toulouse Cedex 4, France}
  M. B\"uttiker,$^b$ and C. W. J. Beenakker$^{a}$
  }

\address{
  $^{a}$Instituut-Lorentz, Leiden University, P.O. Box 9506, 2300 RA
Leiden, The Netherlands\\
  $^{b}$D\'epartement de Physique Th\'eorique, Universit\'e de Gen\`eve,
1211 Gen\`eve 4, Switzerland\\
{}~{\rm (\today)}~ \medskip \\ \parbox{14cm} {\rm
The conductance $G$ of a quantum dot with single-mode ballistic point contacts
depends sensitively on external parameters $X$, such as gate voltage and
magnetic field. We calculate the joint distribution of $G$ and $dG/dX$ by
relating it to the distribution of the Wigner-Smith time-delay matrix of a
chaotic system. The distribution of $dG/dX$ has a singularity at zero and
algebraic tails. While $G$ and $dG/dX$ are correlated, the ratio of $dG/dX$ and
$\sqrt{G(1-G)}$ is independent of $G$. Coulomb interactions change the
distribution of $dG/dX$, by inducing a transition from the grand-canonical to
the canonical ensemble. All these predictions can be tested in semiconductor
microstructures or microwave cavities.
}\bigskip \\ }

\maketitle

\bcols
Parametric fluctuations in quantum systems with a chaotic classical dynamics
are of fundamental importance for the characterization of mesoscopic systems.
The fluctuating dependence of an energy level $E_j(X)$ on an external parameter
$X$, such as the magnetic field, has received considerable attention
\cite{AltshulerSimonsReview}. A key role is played by the ``level velocity''
$dE_j/dX$, describing the response to a small perturbation
\cite{AltshulerSimons1993,Fyodorov1994,Taniguchi}. In open systems, the role of
the level velocity is played by the ``conductance velocity'' $dG/dX$.
Remarkably little is known about its distribution.

The interest in this problem was stimulated by experiments on semiconductor
microstructures known as quantum dots, in which the electron motion is
ballistic and chaotic \cite{Westervelt}. A typical quantum dot is confined by
gate electrodes, and connected to two electron reservoirs by  ballistic point
contacts, through which only a few modes can propagate at the Fermi level. The
parametric dependence of the conductance has been measured by several groups
\cite{Marcus,Chang,Chan}. In the single-mode limit, parametric fluctuations are
of the same order as the average, so that one needs the complete distribution
of $G$ and $dG/dX$ to characterize the system. Knowing the average and variance
is not sufficient. Analytical results are available for point contacts with a
large number of modes
\cite{Jalabert1,Pluhar,Efetov2,Frahm1,FrahmPichard,Rau,Macedo}. In this paper,
we present the complete distribution in the opposite limit of two single-mode
point contacts and show that it differs strikingly from the multi-mode case
considered previously.

\begin{figure}
\vglue -0.99cm
\hspace{0.01\hsize}
\epsfxsize=0.95\hsize
\epsffile{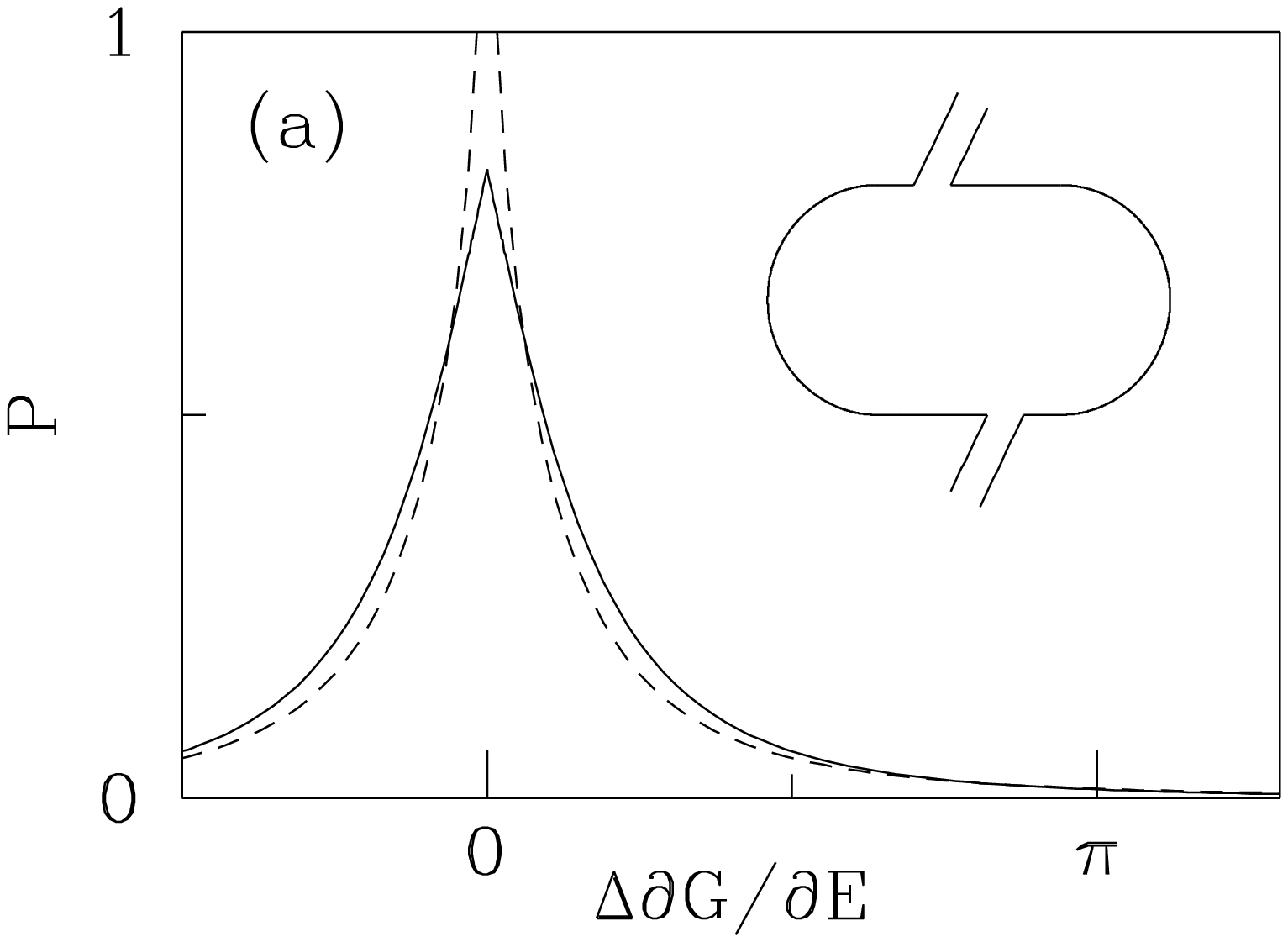}
\vspace{-3.2cm}

\hspace{0.01\hsize}
\epsfxsize=0.95\hsize
\epsffile{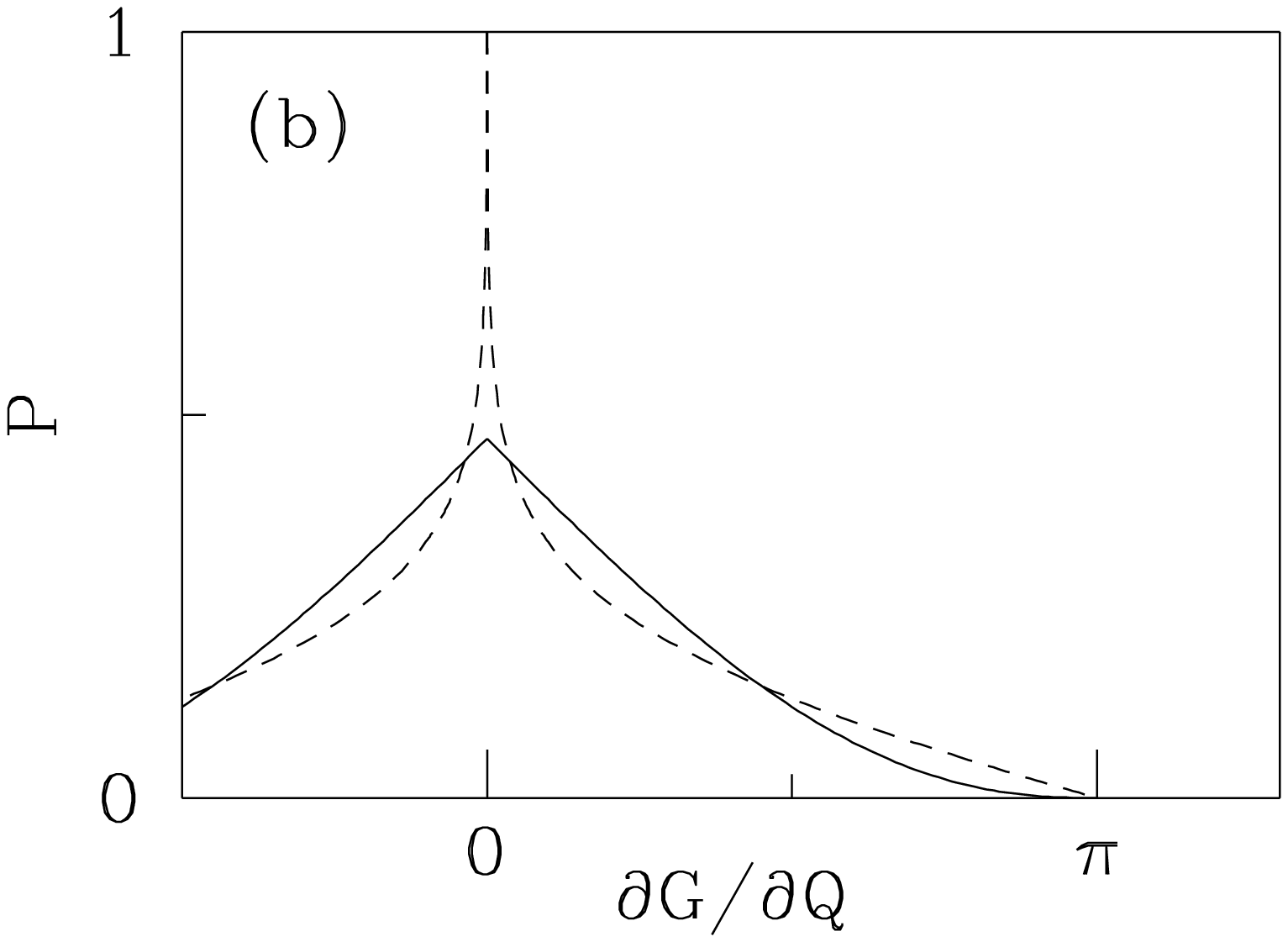}
\vspace{-3.2cm}

\hspace{0.01\hsize}
\epsfxsize=0.95\hsize
\epsffile{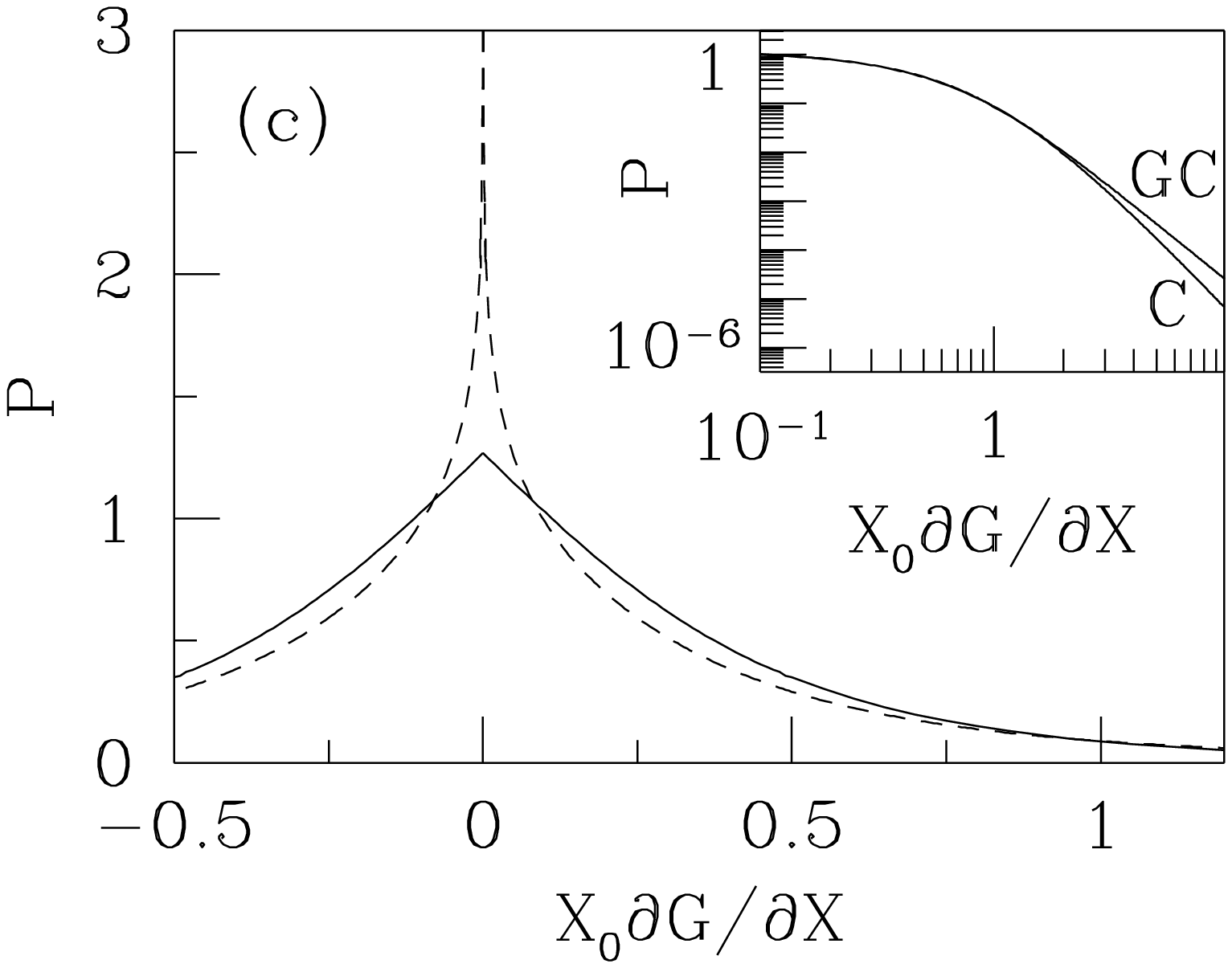}
\refstepcounter{figure} \label{fig:1}

{\small FIG.\ \ref{fig:1}. Distributions of conductance velocities in a chaotic
cavity with two single-mode point contacts [inset in (a)], computed from Eq.\
(\ref{eq:1}). Dashed curves are for $\beta=1$ (time-reversal symmetry), solid
curves for $\beta=2$ (no time-reversal symmetry). (The case $\beta=4$, which is
similar to $\beta=2$, is omitted for clarity.) The distribution of $\Delta
\partial G/\partial E$ (grand-canonical ensemble) is shown in (a) and the
distribution of $\partial G/\partial Q$ (canonical ensemble) is shown in (b).
(The conductance $G$ is measured in units of $2e^2/h$, the charge $Q$ in units
of $e$.) In (c) the distribution of $X_0 \partial G/\partial X$ is shown for
the grand-canonical ensemble (the canonical case being nearly identical on a
linear scale). The inset compares the canonical (C) and grand-canonical (GC)
results for $\beta=2$ on a logarithmic scale. \par}
\end{figure}

The main differences which we have found are the following. We consider the
joint distribution of the conductance $G$ and the derivatives $\partial
G/\partial V$, $\partial G/\partial X$ with respect to the gate voltage $V$ and
an external parameter $X$ (typically the magnetic field). If the point contacts
contain a large number of modes, $P(G,\partial G/\partial V,\partial G/\partial
X)$ factorizes into three independent Gaussian distributions
\cite{Jalabert1,Pluhar,Efetov2,Frahm1}. In the single-mode case, in contrast,
we find that this distribution does not factorize and decays algebraically
rather than exponentially. By integrating out $G$ and one of the two
derivatives, we obtain the conductance velocity distributions $P(\partial
G/\partial V)$ and $P(\partial G/\partial X)$ plotted in Fig.\ \ref{fig:1}.
Both distributions have a singularity at zero velocity, and algebraic tails. A
remarkable prediction of our theory is that the correlations between $G$, on
the one hand, and $\partial G/\partial V$ and $\partial G/\partial X$, on the
other hand, can be transformed away by the change of variables $G = (2e^2/h)
\sin^2 \theta$, where $\theta$ is the polar coordinate introduced in Ref.\
\onlinecite{FrahmPichard}. The derivatives $\partial \theta/\partial V$ and
$\partial \theta/\partial X$ are statistically independent of $\theta$. There
exists no change of variables that transforms away the correlations between
$\partial G/\partial V$ and $\partial G/\partial X$.

Another new feature of the single-mode case concerns the effect of Coulomb
interactions \cite{Buettiker1993,CHR}. In the simplest model, the strength of
the Coulomb repulsion is measured by the ratio of the charging energy $e^2/C$
(with $C$ the capacitance of the quantum dot) and the mean level spacing
$\Delta$. In the regime $e^2/C \gg \Delta$, where most experiments are done,
Coulomb interactions suppress fluctuations of the charge $Q$ on the quantum dot
as a function of $V$ or $X$, at the expense of fluctuations in the electrical
potential $U$. Since the Fermi level $\mu$ in the quantum dot is pinned by the
reservoirs, the kinetic energy $E = \mu - U$ at the Fermi level fluctuates as
well. Fluctuations of $E$ can not be ignored, because the conductance is
determined by $E$, an not by $\mu$. An ensemble of quantum dots with fixed $Q$
and fluctuating $E$ behaves effectively as a canonical ensemble --- rather than
a grand-canonical ensemble. In the opposite regime $e^2/C \ll \Delta$, the
energy $E$ does not fluctuate on the scale of the level spacing. The ensemble
is now truly grand-canonical. Fluctuations of $E$ on the scale of $\Delta$ can
be neglected in the multi-mode case, so that the distinction between canonical
and grand-canonical averages is irrelevant. In the single-mode case the
distinction becomes important. We will see that the distribution of the
conductance velocities is different in the two ensembles. (The distribution of
the conductance itself is the same.) The difference between grand-canonical and
canonical averages has been studied extensively in connection with the problem
of the persistent current \cite{CheungGefenRiedelShih,Bouchiat,Imry}, which is
a thermodynamic property. Here we find a difference in the case of a transport
property, which is more unusual \cite{KamenevGefen,footnote_on_charge}.

To derive these results, we combine a scattering formalism with random-matrix
theory \cite{BeenakkerReview}. The $2 \times 2$ scattering matrix $S$
determines the conductance
\begin{equation}
  G = |S_{12}|^2,
\end{equation}
and the (unscreened) compressibilities \cite{CHR}
\begin{eqnarray}
  {\partial Q \over \partial E} &=& {1 \over 2 \pi i} \mbox{tr}\, S^{\dagger}
{\partial S \over \partial E},\ \
  {\partial Q \over \partial X} = {1 \over 2 \pi i} \mbox{tr}\, S^{\dagger}
{\partial S \over \partial X}.
\end{eqnarray}
(We measure $G$ in units of $2e^2/h$ and $Q$ in units of $e$.) Grand-canonical
averages $\langle \cdots \rangle_{GC}$ and canonical averages $\langle \cdots
\rangle_{C}$ are related by
\begin{equation}
  \langle \cdots \rangle_{C} = \Delta \langle \cdots \times dQ/dE \rangle_{GC}.
\label{eq:NE}
\end{equation}
The factor $dQ/dE$ is the Jacobian to go from an average over $Q$ in the
canonical ensemble to an average over $E$ in the grand-canonical ensemble.
Conductance velocities in the two ensembles are related by
\begin{eqnarray} \label{eq:cangrcan}
  \left.{\partial G \over \partial X} \right|_{Q} &=&
  \left.{\partial G \over \partial X} \right|_{E} -
  {\partial G \over \partial E}
  {\partial Q \over \partial X}
  \left({\partial Q \over \partial E} \right)^{-1},
  \label{eq:cangrcan1}
\end{eqnarray}
where $|_Q$ and $|_E$ indicate, respectively, derivatives at constant $Q$
(canonical) and constant $E$ (grand-canonical). Derivatives $\partial
G/\partial V$ with respect to the gate voltage are proportional to $\partial
G/\partial Q$ in the canonical ensemble and to $\partial G/\partial E$ in the
grand-canonical ensemble. (The proportionality coefficients contain elements of
the capacitance matrix of the quantum dot plus gates.) The two derivatives are
related by
\begin{eqnarray}
  {\partial G \over \partial Q} &=& {\partial G \over \partial E}
  \left({\partial Q \over \partial E} \right)^{-1}.
  \label{eq:cangrcan2}
\end{eqnarray}

The problem that we face is the calculation of the joint distribution of $S$,
$\partial S/\partial E$, and $\partial S/\partial X$. In view of the relations
(\ref{eq:NE})--(\ref{eq:cangrcan2}) it is sufficient to consider the
grand-canonical ensemble. This problem is closely related to the old problem
\cite{WignerSmith} of the distribution of the Wigner-Smith delay times
$\tau_{1},\ldots,\tau_{N}$, which are the eigenvalues of the $N \times N$
matrix $-i S^{\dagger} \partial S/\partial E$. (The eigenvalues are real
positive numbers.) Interest in this problem has revived in connection with
chaotic scattering \cite{FyodorovSommers,Seba,GoparMelloBuettiker,unpublished}.
The rates $\gamma_n = 1/\tau_n$ are distributed according to \cite{unpublished}
\begin{equation} \label{eq:qdistr}
  P(\{\gamma_{n}\}) \propto \prod_{i<j} |\gamma_{i} - \gamma_{j}|^{\beta}
\prod_k
\gamma_{k}^{\beta N/2} e^{-{\pi \beta \gamma_k/\Delta}}.
\end{equation}
This distribution is known in random-matrix theory as the Laguerre ensemble,
because the correlation functions can be written as series over (generalized)
Laguerre polynomials \cite{NagaoSlevin}. For $N=1$ we recover the result of
Refs.\ \onlinecite{FyodorovSommers} and \onlinecite{GoparMelloBuettiker}. In
our case $N=2$.

To compute the conductance velocities it is not sufficient to know the delay
times $\tau_{n}$, but we also need to know the distribution of the eigenvectors
of the time-delay matrix $-i S^{\dagger} \partial S/\partial E$. Furthermore,
we need the distribution of $-i S^{\dagger} \partial S/\partial X$. The general
result containing this information is \cite{unpublished}
\begin{eqnarray}
  && P(S,\tau_{E},\tau_{X}) \propto
  {\exp \left[{- \beta\, \mbox{tr}\,\left({\pi \over \Delta} \tau_E^{-1} +
{\pi^2 X_0^2 \over 4 \Delta^2}(\tau_E^{-1}
\tau_X^{\vphantom{1}})^2\right)}\right]} \nonumber \\
  && \hphantom{P(S,\tau_{E},\tau_{N}) \propto} \mbox{} \times
  {(\det \tau_E)^{-2 \beta N + 3(\beta-2)/2}}, \label{eq:PSQQ} \\
  && \tau_E = -i S^{-1/2} {\partial S \over \partial E} S^{-1/2},\ \tau_X = -i
S^{-1/2} {\partial S \over \partial X} S^{-1/2}. \label{eq:Qdef}
\end{eqnarray}
The matrix $\tau_{E}$ has the same eigenvalues as the time-delay matrix, but it
is more convenient because it is uncorrelated with $S$, while the time-delay
matrix is not. By integrating out $\tau_{E}$ and $\tau_{X}$ from Eq.\
(\ref{eq:PSQQ}), we obtain a uniform distribution for $S$, as expected for a
chaotic cavity \cite{BlumelSmilansky}. The resulting distribution of the
conductance \cite{BarangerMello}, $P(G) \propto G^{-1 + \beta/2}$, is the same
in the canonical and grand-canonical ensembles, because $S$ and $dQ/dE$ are
uncorrelated [cf.\ Eq.\ (\ref{eq:NE})].  By integrating out $S$, $\tau_{X}$,
and the eigenvectors of $\tau_{E}$, we obtain the distribution
(\ref{eq:qdistr}) of the delay times. The distribution of $\tau_{X}$ at fixed
$\tau_{E}$ is a Gaussian. The scale of this Gaussian is set by the parameter
$X_0$, which has no universal value \cite{footnote_on_Xc}.

We are now ready to compute the distribution of the conductance velocities.
Derivatives with respect to $E$ and $Q$ are related to the delay times by
\begin{mathletters}
\begin{eqnarray}
 {\partial G \over \partial E} &=& {c(\tau_1-\tau_2)} \sqrt{G(1-G)},\\
 {\partial G \over \partial Q} &=& 2 \pi c {\tau_1-\tau_2 \over \tau_1+\tau_2}
\sqrt{G(1-G)}, \label{eq:PTheta}
\end{eqnarray}
\end{mathletters}%
where $c \in [-1,1]$ is a number that depends on the phases of the matrix
elements of $S$ and on the eigenvectors of $\tau_E$. Its distribution $P(c)
\propto (1-c^2)^{-1 + \beta/2}$ is independent of $\tau_1$, $\tau_2$, and $G$.
The derivative $\partial G/\partial X$ has a Gaussian distribution at a given
value of $S$ and $\tau_E$, with zero mean and with variance
\begin{eqnarray*}
  \left\langle \left(\left.{\partial G \over \partial X}\right|_{E}\right)^2
\right\rangle &=& \alpha
  \left[ G(1-G) \tau_1 \tau_2 + {1 \over 2} \left({\partial G \over \partial
E}\right)^2 \right],  \\
  \left\langle \left(\left.{\partial G \over \partial X}\right|_{Q}\right)^2
\right\rangle &=& \alpha
  \tau_1 \tau_2 \left[ G(1-G) - {1 \over 4 \pi^2} \left( {\partial G \over
\partial Q} \right)^2 \right],
\end{eqnarray*}
where we have abbreviated $\alpha = 4 \Delta^2/\pi^2 X_0^2 \beta$. Because the
variance of $\partial G/\partial X$ depends on $\partial G/\partial E$ or
$\partial G/\partial Q$, these conductance velocities are correlated.

{}From the distribution (\ref{eq:qdistr}) of $\tau_1$, $\tau_2$, and the
independent distributions of $G$ and $c$, we calculate the joint distribution
of $G$ and its (dimensionless) derivatives $G_X = X_0 \partial G/\partial X$,
$G_E = (\Delta/2\pi) \partial G/\partial E$, and $G_Q = (1/2\pi) \partial
G/\partial Q$. The result in the grand-canonical and canonical ensembles is
\bleq
\begin{mathletters} \label{eq:1}
\begin{eqnarray}
  && P_{GC}(G,G_E,G_X) = {1 \over Z} \int_0^{\infty} dx \int_{{G_E^2 \over
G(1-G)}}^{\infty} dy { \left[y G - G_E^2/(1-G) \right]^{{-1+\beta/2}}
x^{-2-2\beta} \over \sqrt{\pi(x+y)G(1-G) f(x)}} \exp\left [-{2 \beta \over x}
\sqrt{x+y} - {G_X^2 \over f(x)} \right],\\
  && P_{C}(G,G_Q,G_X) =  {2 \over Z} \int_0^{\infty} dx \int_{{G_Q^2 \over
G(1-G)}}^{1} dy { \left[y G - G_Q^2/(1-G) \right]^{{-1+\beta/2}}
  x^{3\beta}
  \over (1-y)^{{(\beta+3)/2}} \sqrt{\pi G(1-G) g(x)}} \exp \left[-{2 \beta x
\over \sqrt{1-y}} - {G_X^2 \over g(x)} \right], \\
  && f(x) = 8 \beta^{-1} [x G(1-G) + 2 G_E^2],\ \ g(x) = 8 (x^2 \beta)^{-1}
[G(1-G) - G_Q^2], \ \ Z = 3 \beta^{-3\beta-1} \Gamma(\beta/2) \Gamma(\beta)
\Gamma(3\beta/2).
\end{eqnarray}
\end{mathletters}%
\eleq

By integrating out $G$ and one of the two derivatives from Eq.\ (\ref{eq:1}),
we obtain the conductance velocity distributions of Fig.\ \ref{fig:1}. (The
case $\beta=4$ is close to $\beta=2$ and is omitted from the plot for clarity.)
The distributions have a singularity at zero derivative: A cusp for $\beta=2$
and $4$, and a logarithmic divergence for $\beta=1$. The tails of the
distributions of $\partial G/\partial X$ are algebraic in both ensembles, but
with a different exponent,
\begin{mathletters}
\begin{eqnarray}
  P_{GC}(\partial G/\partial X) &\propto& (\partial G/\partial X)^{-\beta-2},\\
  P_{C} (\partial G/\partial X) &\propto& (\partial G/\partial X)^{-2\beta-1}.
\end{eqnarray}
\end{mathletters}%
The distribution of $\partial G/\partial E$ (grand-canonical ensemble) also has
an algebraic tail [$\propto (\partial G/\partial E)^{-\beta-2}$], while the
distribution of $\partial G/\partial Q$ (canonical ensemble) is identically
zero for $|\partial G/\partial Q| \ge \pi$. In both ensembles, the second
moment of the conductance velocities is finite for $\beta=2$ and $4$, but
infinite for $\beta=1$.

In conclusion, we have calculated the joint distribution of the conductance $G$
and its parametric derivatives for a chaotic cavity, coupled to electron
reservoirs by two single-mode ballistic point contacts. The distribution is
fundamentally different from the multi-mode case, being highly non-Gaussian and
with correlated derivatives. (Correlations between $G$ and the parametric
derivatives can be transformed away by a change of variables.) We account for
Coulomb interactions by using a canonical ensemble instead of a grand-canonical
ensemble. Our results for the canonical ensemble are relevant for the analysis
of recent experiments on chaotic quantum dots, where the conductance $G$ is
measured as a function of both the magnetic field and the shape of the quantum
dot \cite{Chan}. The grand-canonical results are relevant for experiments on
microwave cavities \cite{microwave1,microwave2}. Together with the theory
provided here, such experiments can yield information on the distribution of
delay times in chaotic scattering that can not be obtained by other means.

This problem was suggested to us by C. M. Marcus. We acknowledge support by the
Dutch Science Foundation NWO/FOM and by the European Community (program for the
Training and Mobility of Researchers).

\ecols

\begin{references}

\bibitem{AltshulerSimonsReview}
        B. L. Altshuler and B. D. Simons,
        in {\em Mesoscopic Quantum Physics}, edited by E. Akkermans,
        G. Montambaux, J.-L. Pichard, and J. Zinn-Justin
        (North-Holland, Amsterdam, 1995).
\bibitem{AltshulerSimons1993} B. D. Simons and B. L. Altshuler,
        Phys.\ Rev.\ B {\bf 48}, 5422 (1993).
\bibitem{Fyodorov1994} Y. V. Fyodorov, Phys.\ Rev.\ Lett.\
        {\bf 73}, 2688 (1994);
        Y. V. Fyodorov and A. D. Mirlin, Phys. Rev. B {\bf 51}, 13403 (1995).
\bibitem{Taniguchi} N. Taniguchi, A. Hashimoto,
        B. D. Simons, and B. L. Altshuler,
        Europhys.\ Lett.\ {\bf 27}, 335 (1994).
\bibitem{Westervelt} R. M. Westervelt, in {\em Nano-Science
        and Technology}, edited by G. Timp (American Institute
        of Physics, New York, 1996).
\bibitem{Marcus} C. M. Marcus, A. J. Rimberg, R. M. Westervelt,
        P. F. Hopkins, and A. C. Gossard, Phys. Rev. Lett. {\bf 69},
        506 (1992).
\bibitem{Chang} A. M. Chang,
        H. U. Baranger, L. N. Pfeiffer, and K. W. West,
        Phys. Rev. Lett. {\bf 73}, 2111 (1994).
\bibitem{Chan} I. H. Chan, R. M. Clarke, C. M. Marcus, K. Campman,
        and A. C. Gossard, Phys.\ Rev.\ Lett.\ {\bf 74}, 3876 (1995).
\bibitem{Jalabert1} R. A. Jalabert, H. U. Baranger, and A. D. Stone,
        Phys. Rev. Lett. {\bf 65}, 2442 (1990).
\bibitem{Pluhar} Z. Pluha\u{r}, H. A. Weidenm\"uller, J. A. Zuk,
        and C. H. Lewenkopf, Phys.\ Rev.\ Lett.\ {\bf 73}, 2115 (1994).
\bibitem{Efetov2} K. B. Efetov, Phys. Rev. Lett. {\bf 74},
        2299 (1995).
\bibitem{Frahm1} K. Frahm, Europhys. Lett. {\bf 30}, 457 (1995).
\bibitem{FrahmPichard} K. Frahm and J.-L. Pichard,
        J. Phys.\ I France {\bf 5}, 877 (1995).
\bibitem{Rau} J. Rau, Phys.\ Rev.\ B {\bf 51}, 7734 (1995).
\bibitem{Macedo} A. M. S. Mac\^edo, Phys.\ Rev.\ B {\bf 53}, 8411 (1996).
\bibitem{Buettiker1993} M. B\"uttiker, J. Phys. Condens.\ Matter {\bf 5},
        9361 (1993).
\bibitem{CHR} M. B\"{u}ttiker and T. Christen, in
        {\em Quantum Transport in Semiconductor Submicron
        Structures}, edited by B.\ Kramer, NATO ASI Series
        {\bf 326} (Kluwer, Dordrecht, 1996).
\bibitem{CheungGefenRiedelShih}
        H.-F. Cheung, Y. Gefen, E. K. Riedel, and W.-H. Shih,
        Phys.\ Rev.\ B {\bf 37}, 6050 (1988).
\bibitem{Bouchiat} H. Bouchiat and
        G. Montambaux, J. Phys.\ (Paris) {\bf 50}, 2695 (1989).
\bibitem{Imry} B. L. Altshuler, Y. Gefen, and Y. Imry, Phys.\
        Rev.\ Lett.\ {\bf 66}, 88 (1991).
\bibitem{KamenevGefen} A. Kamenev and Y. Gefen, Phys.\ Rev.\ B {\bf 49},
        14474 (1994).
\bibitem{footnote_on_charge}
        The difference between canonical and grand-canonical
        averages which we find is related, but not identical, to
        the effects of the Coulomb blockade predicted by I.\ L.\
        Aleiner and L.\ I.\ Glazman (preprint, cond-mat/9612138).
\bibitem{BeenakkerReview} C. W. J. Beenakker,
        Rev.\ Mod.\ Phys.\ (April 1997, cond-mat/9612179).
\bibitem{WignerSmith} E. P. Wigner, Phys.\ Rev.\ {\bf 98}, 145 (1955);
        F. T. Smith, Phys.\ Rev.\ {\bf 118}, 349 (1960).
\bibitem{FyodorovSommers} Y. V. Fyodorov and H.-J. Sommers,
        Phys.\ Rev.\ Lett.\ {\bf 76}, 4709 (1996);
        J. Math.\ Phys.\ (to be published, cond-mat/9701037).
\bibitem{Seba} P. \v{S}eba, K. {\.Z}yczkowski, and J. Zakrewski,
        Phys. Rev. E {\bf 54}, 2438 (1996).
\bibitem{GoparMelloBuettiker} V. A. Gopar, P. A. Mello, and
        M. B\"uttiker, Phys.\ Rev.\ Lett.\ {\bf 77}, 3005 (1996).
\bibitem{unpublished} P. W. Brouwer, K. M. Frahm, and C. W. J. Beenakker
        (unpublished).
\bibitem{NagaoSlevin} T. Nagao and K. Slevin, J. Math.\ Phys.\
        {\bf 34}, 2075 (1993); T. Nagao and P. J. Forrester,
        Nucl.\ Phys.\ B {\bf 435}, 401 (1995).
\bibitem{BlumelSmilansky} R. Bl\"umel and U. Smilansky, Phys. Rev. Lett.
        {\bf 60}, 477 (1988).
\bibitem{BarangerMello} H. U. Baranger and P. A. Mello,
        Phys.\ Rev.\ Lett.\ {\bf 73}, 142 (1994);
        R. A. Jalabert, J.-L. Pichard, and C. W. J.
        Beenakker, Europhys.\ Lett.\ {\bf 27}, 255 (1994).
\bibitem{footnote_on_Xc}
        If $X$ represents the magnetic flux through the quantum dot,
        then $X_0 \simeq (h/e) (\tau_{\rm ergodic}/
       \tau_{\rm dwell})^{1/2}$, where $\tau_{\rm dwell}$ is
        the mean dwell time in the quantum dot and
        $\tau_{\rm ergodic} \ll \tau_{\rm dwell}$ is the time
        scale for ergodic exploration of the available phase space.
\bibitem{microwave1} J. Stein, H.-J. St\"ockmann, and U. Stoffregen,
        Phys.\ Rev.\ Lett.\ {\bf 75}, 53 (1995).
\bibitem{microwave2} A Kudrolli,
        V. Kidambi, and S. Sridhar, Phys.\ Rev.\ Lett.\ {\bf 75},
        822 (1995).

\end{references}
\end{document}